\documentclass[AMA,STIX1COL]{WileyNJD-v2}
\usepackage{moreverb}

\usepackage[utf8]{inputenc}
\usepackage{todonotes}

\articletype{Article Type}%

\received{<day> <Month>, <year>}
\revised{<day> <Month>, <year>}
\accepted{<day> <Month>, <year>}

\begin{document}

\title{Changes in Conducting Data Protection Risk Assessment: Before and After GDPR implementation}
\author[1]{Fatemeh Zarrabi}
\author[1]{Isabel Wagner}
\author[1]{Eerke Boiten}

\address[1]{\orgdiv{Cyber Technology Institute}, \orgname{De Montfort University}, \orgaddress{\state{Leicester}, \country{UK}}}

\corres{*Corresponding author name, Corresponding address. \email{authorone@Email.com}}

\abstract{Based on Article 35 of the EU (European Union) General Data Protection Regulation, a Data Protection Impact Assessment (DPIA) is necessary whenever there is a possibility of a high privacy and data protection risk to individuals caused by a new project under development.  A similar process to DPIA had been previously known as Privacy Impact Assessment (PIA). We are investigating here to find out if GDPR and DPIA specifically as its privacy risk assessment tool have resolved the challenges privacy practitioners were previously facing in implementing PIA. To do so, our methodology is based on comparison and thematic analysis on two sets of focus groups we held with privacy professionals back in January 2018 (four months before GDPR came into effect) and then in November 2019 (18 months after GDPR implementation).}


\maketitle

\section{Introduction}
Early treatises on privacy appeared with the development of privacy protection in American law from the 1890s onward \cite{warren1890right}. They defined privacy as the right of individuals to be left alone. Therefore, one can accord privacy as a value to control the access others have to him or her \cite{warren1890right}. The right to privacy has later been defined in different aspects such as in individual's physical places, behaviors, their communication and information and their decisions \cite{birnhack2008eu,clarke1997introduction,ICO}. In contrast, data protection is a more recent terminology in legislation to control how personal data are being used in organizations, business or governments \cite{ICO-GDPR}. Privacy and data protection are two concepts that these days are sometimes used interchangeably, erroneously. Privacy is about all aspects of human dignity including individual integrity, personal autonomy, anonymity, confidentiality, independence, etc.\  \cite{wright2014privacy} and that is why a threat to privacy is a threat to our very integrity as natural persons. Therefore, considering all types of privacy rights and principles is the proper avenue to identifying most possible harms consequent from non-compliance to privacy and to taking a more proper approach to privacy risk assessment \cite{wright2014privacy}. 

Collecting an individual's personal data and its subsequent processing gradually transfers control from the data subject to the data controller. There are several perspectives of privacy rights in information. One follows from data being highly valued in the market, and in that spirit adapts a model of ownership of data. In that view, the main owner of an individual's personal data is not the data subject her/himself, but the data controller. Therefore, the responsibility of any failure addresses the market owner (data controller) and consequently harms against the owner's interest will be the first concern in this case. In contrast, another perspective focuses on the data subject and personal data is understood in terms of human rights with a proprietary tone \cite{birnhack2008eu}.  The effects and harms of abuse and non-adherence to privacy principles are significant, multidimensional, and sometimes irrecoverable. It will be vital not to limit the harms of violation to privacy in data protection only to business damages. Hence, precise attention should consider all types of harms including the ones on personal dignity. Therefore, it will be essential and especially important to first estimate accurately the possibility of any detriment to privacy, and secondly to assess all types of harms and level of affects they may cause to the data subject and any additional stakeholders if such harms occur. This process forms the core of Data Protection Impact Assessment (DPIA) within the General Data Protection Regulation (GDPR).

\subsection{Problem statement}
How to conduct a systematic DPIA is an open question and challenge for both academic researchers and privacy professionals \cite{ferra2020challenges,binns2017data}. Open issues include, for example, how different types of harms should be considered, how risk probabilities and impacts can be measured quantitatively, how individual versus organizational impacts should be considered, and how effective current frameworks and guidelines are. 
Researchers have highlighted the need for a generic framework for privacy risk measurement and the importance of streamlining privacy risk assessment processes
\cite{de2012human,ferra2020challenges,wright2014integrating}.
In addressing these issues, however, it is important to take into account the perspective of privacy professionals, including their experiences, daily activities, and current practices in conducting privacy risk assessments.

\subsection{Contribution}
In this paper, we analyze the privacy risk assessment practices of a group of privacy professionals which we recorded during two focus groups held before and after the GDPR came into force.
We compare practices before and after the GDPR to find out whether the GDPR has addressed the challenges of conducting pre-GDPR Privacy Impact Assessments (PIAs), and whether the new Data Protection Risk Assessments (DPIAs) have introduced new challenges.
%
Our main findings are: 

\begin{itemize}
\item Defining Privacy: Understanding of privacy definition has not improved, and there is no widely accepted definition of privacy yet. Instead definitions has shifted from privacy to data protection.
\item Awareness: GDPR has had a global impact in increasing awareness  regarding to data protection, but this impacts mainly on the necessity of privacy protection and not so much on related tasks and responsibilities. for example, users of some very well-known social media services are not still aware of privacy control tasks and options on these services or how to claim their privacy rights.
 \item Education: It has become mandatory to educate organizations' staffs about data protection, but the level of training tends to be weak.  
 \item Digital Era: There is improvement of awareness about privacy challenges of recent technologies, however despite GDPR enforcement modern technologies still have privacy issues and DPIA is not being implemented in most recent digital technologies. 
  \item Human Rights: Despite data subjects' rights provided by GDPR, individuals still do not have enough control over their data sharing in some applications, and their choices are not always fully informed. 
 \end{itemize}

Remaining challenges for DPIA practitioners that emerged from the focus groups were the following.

\begin{itemize}
    \item DPIAs procedure: The process is difficult to run because comprehensive guidelines or platforms are either unavailable or perceived as not very useful in practice. 
    \item DPIA Guidelines: Manual and templates are useful mainly to demonstrate compliance, but less useful for substantive consideration and implementation of data protection.
    \item Expertise: Privacy professionals are typically not trained risk assessors, so a lack of expertise in risk assessment has caused difficulty in running DPIAs.  
    \item Measurement: Likelihood and impact are hard to be measured, with impact assessment perceived as easier than likelihood assessment. 
    \item Privacy by design: The practical meaning of concept is unclear, but it has already been practiced by most. As it is said, it is like an elephant in the dark room.
    \item DPIA and assessment of project risk: No tool to integrate both. Although some practitioners argue for separation of these tasks, but GDPR suggests integration of DPIA with other organizational risk assessment processes. 
    \item Transparency: There is a lack of transparency around DPIA processes, particularly as it is not required to publish DPIAs' outcomes.
    \item Compliance: Companies claim 100\% compliance to GDPR, but provide only weak evidence such as ``we have trained staffs''.
    \item Ethic: Regulatory compliance is not equal to ethical compliance or compliance to privacy in human rights terms.
    \item Roles: Unclear which team in companies should fulfill the data protection officer role or which team should run DPIA.
    \end{itemize}

The remainder of this paper is structured as follows. We introduce related work in Section \ref{sec:background} and discuss our methodology in Section \ref{sec:methodology}. In Section \ref{sec:Thematic}, we present the results of our thematic analysis and comparative analysis along with the identified themes and codes. We discuss our findings in Section \ref{sec:Discussion}, and Section \ref{sec:Conclusion} concludes the paper.

\section{Background}
\label{sec:background}
Data protection frameworks initially came to implementation by guidelines such as the ``Recommendations of the Council Concerning Guidelines Governing the Protection of Privacy and Trans-Border Flows of Personal Data'' \cite{horodyski20142013} by the OECD in 1980, which became the basis of many national laws for data protection. To make a unique law for data protection in Europe and resolve disparities and obstacles in transferring personal data across national territories, the European Parliament approved the Directive 95/46/EC in 1995. Among many benefits of the Directive such as harmonization in the Europe for data protection laws and enormous impact on the globe \cite{bennett1990formation}, it also has been targeted by many critiques such as lacking enforcement of the law, the need for more clarification \cite{birnhack2008eu}, challenges in transferring personal data due to differences in national implementations, and the application of Directive on broad dimensions of internet activities in the online world \cite{bu2017cross}. During the first decade of the 20th century, Data Protection Directive 95/46/EC became the subject of wide-ranging reviews to make it more effective in response to the growth in the field of information technology and the ways it would hugely affect the processing of personal data \cite{hustinx2013eu}. Research for the UK Information Commissioner's Office (ICO) in 2008 on the strength and weaknesses of Directive 95/46/EC \cite{robinson2009review} suggested the Directive as one of laws to be reviewed by the Better Regulation Agenda. After many different consultations with diverse groups such as citizens, businesses, governmental bodies, Data Protection authorities, etc by the European Commission, they finally decided to modernize the legal framework with a regulation called the General Data Protection Regulation (GDPR). After the extensive EU legislative process, GDPR came into force in May 2018. One of the main challenges of implementing GDPR in the first years after its introduction has been for companies to understand what changes it imposed \cite{tikkinen2018eu}. In addition, organizations faced difficulties in understanding compliance to the GDPR requirements \cite{sirur2018we,agarwal2016towards,tankard2016gdpr}. In response, sector bodies supplied guidelines on implementation of the GDPR \cite{team2020eu}. 

Privacy Impact Assessment (PIA), as an instrument for data protection regulatory regime, has been developed for a long time before the wide scale use of the internet, big data, artificial intelligence and mobile technology \cite{raab2020information}. The early development and implementation of PIA had been in countries such as North America, Canada, New Zealand and Australia \cite{weaver2013corporate,robinson2009review}. The EU Data Protection Directive 95 did not directly foresee the use of PIA.  Also, as there was no obligation for such a process and due to cultural differences, attempts to prescribe PIA in practice in different member states led to a wide range of approaches ( e.g.\ the French (CNIL) PIA Methodology \cite{CNIL}, BfDI in Germany \cite{wright2013comparative} and the UK  (ICO) Handbook on PIA \cite{ICO}). In 2007, a research team in Loughborough University were commissioned by the UK ICO to study PIA guidance and practices in jurisdictions with PIA high standards such as Canada, New Zealand and Australia and to identify learned lessons for a PIA handbook in the UK \cite{wright2013comparative}. Several critiques on the UK PIA process and the ICO handbook were issued after. Warren and Charlesworth \cite{warren2012privacy} identified problems such as lack of review and oversight on PIA process and a lack of PIA cross-fertilization across departmental boundaries. Research commissioned by the European Commission in 2011 under the name of PIAF (Privacy Impact Assessment Framework) reviewed all PIA methodologies from the most advanced countries in this domain \cite{warren2012privacy}. Many member states responded to a survey \cite{wright2013making} with a preference for a streamlined, quick, easy-to-understand and easy-to-use PIA methodology. All these views contributed to the eventual introduction of Article 35 of GDPR to mandate ``Data Protection Impact Assessment (DPIA)'' as a method to assess high risks of personal data processing against rights of data subjects \cite{EU}. A study in 2012 \cite{de2012human}, tried to identify differences between PIA and DPIA from a human rights point of view. Different supervisory authorities and individual organizations have set up guidelines for DPIA, but so far there is no comprehensive view of these guidelines. Binns \cite{binns2017data} argued PIA have shifted from a self-regulation and voluntary tool into a meta-regulation tool as DPIA.  Another study in the Netherlands \cite{van2017privacy} ran interviews with data protection officers and concluded that business impacts are taken as focal point instead of data subjects' rights in DPIA procedures. 

As part of research into the quantification of privacy risk \cite{wagner2018privacy}, the authors held focus groups in January 2018 with data protection practitioners. This sought to extract knowledge on how professionals assess privacy risk in (D)PIA processes. The outcomes of this workshop have been reported in \cite{ferra2020challenges} and covered a wide range of insights into data protection impact assessment processes in practice. As this took place before the start of GDPR, with DPIA less embedded in organizational processes, we repeated focus groups with the same questions in November 2019. 

This paper reports on the outcomes of these focus groups, in terms of the professionals' views on effective DPIA, as well as a comparison of the positions 4 months before vs.\ 18 months after the start of enforcement of the GDPR.

\section{Methodology}
\label{sec:methodology}
This work is based on data from two sets of focus groups with privacy professionals.
The first focus groups were held in January 2018, a few months before the GDPR came into force. The focus groups were part of a one-day workshop, attended by 16 external privacy professionals. 
We transcribed the recorded conversations and analyzed them using inductive thematic analysis. This analysis showed that measurement of privacy risks and conducting PIA was a difficult procedure which would need more academic attention and research \cite{ferra2020challenges}.

The second focus groups were held in November 2019, 16 months after the GDPR came into force. The participants of the second focus groups partially overlapped with the first group and included privacy consultants, data protection officers, and academics (independent from the authors), see Table \ref{tab:participants}. We again transcribed the recorded conversations.

To analyze the transcripts from the second set of focus groups, we first used inductive Constant Comparison Analysis \cite{leech2007array} (Thematic Analysis) to identify new codes and themes. We then used deductive Constant Comparison Analysis \cite{leech2007array,onwuegbuzie2009qualitative} to compare the themes and codes from the first focus groups with the new themes. 
\begin{table}[ht]
    \centering
    \begin{tabular}{ll}
    \toprule
    Expertise & Participant labels \\
    \midrule
        Professional privacy consultant & CON1, CON2, CON3, CON4, CON5, CON6 \\
        Data protection officer & DPO1, DPO2 \\
        Academic &  ACA1, ACA2 \\
        \bottomrule
    \end{tabular}
    \caption{Expertise of participants in our focus groups and plenary session}
    \label{tab:participants}
\end{table}
%

\section{Results from the Thematic analysis and Comparison Analysis}\label{sec:Thematic}
In our earlier study, we analyzed the first of the two focus groups and identified three main themes \cite{ferra2020challenges}: privacy in contemporary society and how this affects privacy risk assessment, current practices and procedures in privacy risk assessment, and issues \& challenges. 
Here, we focus on how these themes have changed after introduction of the GDPR.
We observe that some of the original themes are still present and nearly unchanged, some have disappeared, some have expanded or changed significantly, and some new themes have emerged.
Table 2 compares the changes in themes between the first and second focus groups, i.e., between the pre-GDPR and post-GDPR situation.

\begin{table}[ht]
    \centering
    \begin{tabular}{p{4cm}lll}
    \toprule
    Theme & Sub-theme & pre-GDPR & post-GDPR \\
    \midrule
        \multirow{3}{4cm}{Privacy in contemporary society and how this affects privacy risk assessment} & Defining Privacy & $\checkmark$ & changed \\
        & Digital Era & $\checkmark$ & changed \\
        & Ethics, human rights, and the law & $\checkmark$ & changed \\
    \midrule
        \multirow{15}{4cm}{Current practices and procedures in privacy risk assessment} & Quantifying privacy impact & $\checkmark$ & unchanged \\
        & Content and context & $\checkmark$ & changed \\
        & Process rather than template & $\checkmark$ & -- \\
        & Types of impact & $\checkmark$ & unchanged \\
        & Impact vs probability & $\checkmark$ & changed \\
        & Identifying high-risk projects & -- & new \\
        & Tools and guidance for DPIA & -- & new  \\
        & Treatment Controls & -- & new \\
        & Integration of DPIA with Project Risk Assessment & -- & new \\
        & Training & -- & new \\
        & DPIA Transparency & -- & new \\
    \midrule
        \multirow{6}{4cm}{Issues \& challenges} & Over-complicated and time-consuming & $\checkmark$ & changed \\
        & It is all about guts & $\checkmark$ & -- \\
        & Moving fast & $\checkmark$ & -- \\
         & Compliance with the GDPR & -- & new \\
        & Privacy Roles & -- & new \\

        
    \bottomrule
    \end{tabular}
    \caption{Comparison of themes identified in pre-GDPR and post-GDPR focus groups.}
    \label{tab:codes}
\end{table}

\subsection{Privacy in contemporary society and how this affects privacy risk assessment}

\subsubsection{Defining Privacy}

GDPR is believed to be a data protection landmark with its principles based on privacy and human right concepts through a privacy evolutionary in Europe \cite{goddard2017eu}. Therefore, it has been expected to increase public and professional understanding of privacy concepts, principles and definition. However, data gathered from our practitioner's conversation in the second study does not prove it as it can be found below. The other concern is also the shift from privacy concepts to data protection after GDPR, although our practitioners believe on the difference between these twos and the fact that DPIA does not address privacy expectation of individuals in most cases.  

\begin{quote}
\textit{a lot of people don't even really understand the definition of personal data, which is not just something that could identify, it's something from which somebody could be identified.} (DPO1)
\end{quote}

\begin{quote}
\textit{there's two different things for me. One is ...privacy and the other side is data protection and the two are actually not the same...I'm not necessarily there to protect the privacy of the individuals, two separate things...Quite often, the way that privacy has been perceived as being impacted is probably not project that requires DPIA in the first place.} (CON1)
\end{quote}

Compared to the pre-GDPR study, definitions and conceptualization are now more driven from GDPR terminologies, resulting in a shift from \textit{privacy} to \textit{data protection}. This is also evident from the text of the GDPR where the first article defines the regulation's objective as the protection of natural persons with regards to the processing of personal data \cite{EU}. In contrast, the first article of the GDPR's predecessor, the Data Protection Directive, defined the objective to protect fundamental rights of natural persons, in particular to privacy, with respect to processing of personal data \cite{birnhack2008eu}. This is concerning because privacy has much broader application than data protection regarding human rights of individuals. 

\subsubsection{Digital era}

Article 22 of the GDPR (Art. 22) \cite{EU} prohibits automated decision making and profiling, with a small number of exceptions including necessity of contract and explicit consent.
However, different types of digital era technologies and applications from educational monitoring software in schools to passport monitoring applications in airports, Facebook and black data markets have been called by our participants and discussed as areas of concern not only for privacy, but also regarding to other ethical and moral matters such as bias and diversity. The major concern had been discussed as the lack of privacy and ethical practices in schools and other authorities, such as not a pre-given and informed consent to pupil's parents, or lack of DPIA process or compliance to ISO27001 and other relevant standards in schools. 


\begin{quote}
\textit{Pupils, 11 year old children, are still prepared to give up their privacy in order to use networking sites such as Snapchat and Instagram and teachers are prepared to give that up for the convenience of doing some assessment tracking.} (CON3)
\end{quote}

\begin{quote}
\textit{most people will compromise their privacy for convenience or for perceived benefit} (DPO1)
\end{quote}

\begin{quote}
\textit{There's some software that will track children's mental well-being within the school...  alarms will be sent if, a child looks at mental health things on the internet, it's all monitored, and I think it's really alarming.} (CON3)
\end{quote}



Raising and promoting awareness of public, data controllers, processors and organizational staffs about privacy risks, rules, safeguards and rights have been articled (ART. 57 and ART.39) in GDPR \cite{EU} as the necessary tasks of supervisory authorities and data protection officers. The global impact of GDPR and its effectiveness in increasing public and professional's awareness has been mentioned by many authors recently \cite{goddard2017eu,kuner2012european,tikkinen2018eu} and is also evident in our post-GDPR focus group. However, although the level of awareness and education has been reported to have improved post-GDPR, it appears to have progress only on the necessity of subject, and not on tasks and responsibilities for personal data processing. 

\begin{quote}
\textit{On a positive note, the awareness has really improved.  Even if people don't understand the responsibility, they understand at least that it is necessary, by and large.} (DPO1)
\end{quote}

\begin{quote}
\textit{We've got a teacher who shows a parent every child's record...what we can do as a group to raise awareness?} (DPO2)
\end{quote}

It was also reported that people were unprepared and struggling to take safeguarding steps protecting their privacy due to limited educational provision on privacy and privacy risks. Privacy practitioners are worried as individuals are not aware of the way these technologies work, giving them their personal data to achieve for other benefits.  

\begin{quote}
\textit{Yeah, they might not be totally aware of what they're signing up for.} (CON5)
\end{quote}

\begin{quote}
\textit{I think people will trade privacy for convenience.} (DPO2)
\end{quote}

Despite the GDPR regime and improvement in level of awareness about privacy issues and data protection to become an interesting and well-known subject among both public and practitioners, however it is still poor in implementation.

\subsubsection{Ethics, human rights, and the law}

The willingness, rights and control of people on the process being held on their personal data if it is a local process or sharing with third parties, have being hugely practiced as one of the focuses of GDPR through number of principles and articles. Article 6 of GDPR has precisely obligated the requirements for consent of data subject as one of the main controlling tools on processing personal data. Providing privacy notice to data subjects as a method practice of transparency has been clearly defined as an obligation of controllers in Articles 12, 13 and 14 along with many other controlling and access rights of data subjects   to their personal data in Articles 15 to 22. However, article 23 restricts these rights when it is necessary for national security, defense or public security. Article 58 of GDPR, gives powers to supervisory authorities (e.g., ICO) to investigate the performance of above rights of data subjects on the practices of controllers and processors under their authorities. Despite all these regulatory rules, it was still noted in our study that individuals did not have enough control on their data sharing in some domains, and it is not always an informed choice. It depends on companies manager's perception of people's rights, limiting the extent on human rights and ethics.  

\begin{quote}
    \textit{Well look at Facebook, they were fined five-billion, that’s nothing to them, I mean it's ridiculous. [...] I would like one of these European regulators to invoke Article 58.2F\footnote{Article 58.2F states that supervisory authorities have the power to impose a ban on processing.} to put a temporary, or possibly permanent, limit on Facebook processing data, and see what happens.  Because fining them doesn't do anything, they've got money, but if you tell them you can't process data, then that becomes another story.} (CON6)
\end{quote}

\begin{quote}
\textit{By you slamming Article 58.2F you're basically taking away my right, because I should be the one who should be making decisions, whether I want to give my data to Facebook or not.} (CON4)
\end{quote}



Human right is still optional for business owners and public still do not know how to claim their rights such as to have control over their personal data processing.
Whereas in our pre-GDPR study, participants discussed the lack of a comprehensive legislation, this discussion has shifted in the post-GDPR study to how the new legislation is, can, and should be, applied.

GDPR also wants the data controllers to seek the view of data subjects or their representatives, whenever appropriate, on the intended processing of their personal data. This is to perceive their expectations, concerns and needs regarding to their privacy when the processing is in development stages. Our practitioners also addressed the need for these communication and perceptions of privacy expectations from data subjects as soon as possible in design of processing systems and with quick, understandable and easy questions in format of questionnaires. They found lack of the practical measures for this important. Also the fact that people with different background have different views and expectation of privacy, makes it something rational. Doing a citizen's science project was suggested as a method to identify public expectations and experiences on personal data processing.

\begin{quote}
\textit{We need the time to help people empathize with data subjects as far as they're able and explore real life risk scenarios in their own risk context versus the better answer to business innovational marketing kind of benefits, or risks.} (CON2)
\end{quote}

\subsection{Current practices and procedures of DPIAs}

\subsubsection{Quantifying privacy impact}

DPIA as a risk assessment tool developed under GDPR to manage the risks of personal data processing (we say privacy!), needs to evaluate and measure some factors associated with risk management and measurement. These factors as being addressed by GDPR and identified in our studies are two major elements in measuring any types of risks known as impact (severity), and the likelihood (probability). There are two main approaches to risk assessment: qualitative and quantitative, which the later has been more favored by risk assessment community for being more tangible and measurable \cite{aven2018society}. The procedure of DPIA along with the methods how the two mentioned factors of risk assessment are measured in DPIA, has been discussed in detail in our new study. The approaches taken by our participants look to be more semi-quantified. Participant addressed difficulty of having a wholly quantified approach mostly because of the difficulty to quantify the level of likelihood and impact. The approaches with more quantifying measurement such as NHS incident response matrix have been reported as better and more useful.

\begin{quote}
\textit{I'm really interested in some of the ...pseudo-quantitative,... things like factor analysis of information risk, } (CON2)
\end{quote}

\begin{quote}
\textit{... quantitative assessment of privacy, related impact and instant probability is an aspiration, } (CON2)   
\end{quote} 

Article 35 of GDPR, has made it mandatory for controllers to conduct a data privacy impact assessment prior to the processing on projects with high risks on rights and freedom of protection of individuals personal data \cite{EU}. Controllers are to seek advice from data protection officer to run this process and supervisory authorities are obligated to provide publicly available guidelines and template to when and how to do DPIA. Despite all these efforts, there was an agreement between our newly interviewed practitioners for the difficulty of running DPIA and the necessity of an easier and simpler approach for DPIA.  Generally Performing impact or risk assessment has been reported as a complex task in the past in different research \cite{OECD,meis2015supporting}  as it is believed that assessing impact and risk are problematic by their nature. 

\begin{quote}
\textit{How do you assess privacy risk, with difficulty I think is the answer.} (CON1)
\end{quote}

\begin{quote}
\textit{We are still, as we were in January '18, in quite a position of making an educated guess based on our knowledge and experience.} (DPO1)
\end{quote}

Main tasks such as likelihood and impact measurement have remained challenging and mostly based on guess rather than scientific and quantifying approaches. 

\subsubsection{Content and context}

One of the main obstacles in performing DPIA in different organizations, has been mentioned as the dependency of DPIA and privacy risk to the context. It is again due to the nature of risk and its dependency to many factors which are known as risk context and criteria. These elements are identified and assessed in initial stages of risk assessment in well-known risk management approaches such as ISO 27005
\cite{schweizerische2013information}. Non-qualified risk assessors may face with difficulty in running DPIA, if they are not familiar with standard procedures of risk assessment. Conducting DPIA by a qualified risk assessor or adopting a systematic and research-based risk assessment approach specifically for DPIA, instead of borrowing already available risk assessment approaches that are designed and used in other domains, look to be solution to many of current issues in running DPIA.

\begin{quote}
\textit{The thing about risk, unless you're a qualified risk manager, then everybody has slightly different ways of doing it.} (CON1)
\end{quote}

The first study identified the role context plays in determining the sensitivity of data and the influence of changes over time. In contrast, in the second study the focus shifted to highlight the role of the risk assessor's experience in determining privacy risks.

\subsubsection{Types of impact}
We had long and interesting conversations in our focus groups about factors in measuring impact.  For example, it was stressed how organizations (mostly commercials) may not consider impacts by regulators such as ICO very serious because they had been advised that ICO fines will always be proportionate. But instead, their priority is on direct business impact (such as fines, customer loss, etc.). Indirect impacts such as customers retention or data restriction, net promoter score, market share,class action and focus on organizational impact rather than impact on individuals were among major concerns  addressed. A massive historical and even legislative gap in impact analysis has been reported  as lack of understanding of harms on individuals and considering only physical or financial harms as tangible impacts. However, some participants disagreed with this opinion, claiming that they already are using approaches for personal impacts measurements such as Enisa. In another conversation, different factors that play important roles on impact measurement have been discussed such as size of data and its sensitivity, context (country, type of organization, culture, etc.) and user expectation.  For example, it was discussed how a same incident on data subject's data in the same context, may have different impacts on different groups of individuals from different ethnicity background or a special reputation or sensitivity. Sectioning impacts on different groups of individuals in DPIA process or assessing user's expectation, have been mentioned as solutions to this problem. 

\begin{quote}
\textit{Real quantifiable financial impacts, psychological impacts, and then collective impacts, which is the thing which is most often missing, societal impacts, so impacts on discreet groups of society, from incidents.} (CON2)
\end{quote}

\begin{quote}
\textit{We have a focus on historic physical harm and financial damage in the virtual spaces, we really have a judicial and legislative gap of understanding the harm to your persona as you are on the internet.} (CON2)
\end{quote}

\begin{quote}
\textit{how sensitive is it, are we special category data, are we children, are we criminal records,...are we dealing with vulnerable groups of people...  they may be groups that people have an interest in doing harm to collectively and they single out, or as individual people they may be far more vulnerable to harm because of their circumstances than other people.} (CON2)
\end{quote}

 \begin{quote}
\textit{the sectioning up the kinds of impacts that data subjects are suffering to bring a bit more perspective on that in a more consistent way to people.} (CON2)
\end{quote}

\begin{quote}
\textit{...you've got a group of people in a room, some people in that room might not want to ever let it be known that they've, let's say, had a heart attack, but there will be others in the room...who don't care } (CON1)
\end{quote}

\subsubsection{Impact vs probability}

It was overall agreed by our participants that measuring level of impact is a much easier task than assessing likelihood, because impact can be quantified to some extent. But likelihood is more a guess than a tangible fact. Need for a centralized intelligence from different organizations about their annual data breaches has been discussed as an essential for a more precise likelihood assessment. In the absence of such a central repository, good communication, training and sharing data in organizations have been suggested as an alternative.  Some participants even found the whole process of likelihood assessment difficult for them as they are not qualified risk assessors. But others could address some factors in probability measurements such as the level of impact (more financial gain more chance of attack), type and sensitivity of data, type of individuals being involved (celebrities), or security controls being taken, easiness of the attack or disclosure and the history log statistic data and historical incidents.

 \begin{quote}
\textit{it's still a level of finger in the wind in my opinion.} (CON1)
\end{quote}

\begin{quote}
\textit{You need to get industry-wide movements to share their risk information.  We need to get some centralized oomph behind that happening is my perspective.} (CON2)
\end{quote}

\begin{quote}
\textit{struggle with likelihood, and that might be the fact that I am not a risk manager} (CON1)
\end{quote}

\begin{quote}
\textit{So I was thinking again from technical point of view, if there's no encryption, if there is no authorization to access the data, I would say the likelihood's going to be very high.} (ACA1)
\end{quote}

Compared to the pre-GDPR study, the idea of a centralized repository of likelihood information, similar to databases that cyber risk insurers rely on, is a new aspect.

DPIA procedures are still difficult or not clear to run.

\subsubsection{Identifying high-risk projects} 
According to GDPR, data controllers are obligated to run a DPIA on projects that are likely to result into high risks to data protection, and not for all projects. But how to identify highly risks projects is an open question to privacy practitioners. Article 29 Working Party has created a catalog of ten criteria indicating processes with high risk \cite{EC-Data-protection}. Article 35 of GDPR also has made it mandatory for supervisory authorities to publish a list of processing with high risk (and with no high risk) which demand DPIA to the public. Our interviewees also addressed some of measures which can be used to identify highly risk projects based on available guidance and platforms for DPIA, or based on their experiences. However they still found the analysis to identify highly risks project not easy.

\begin{quote}
\textit{the main thing that came out of that was a list of things that they defined as predisposing processing to high risk.  So there are a range of things that they grouped into stuff like aggregate processing, or biometric data use.} (CON2)
\end{quote}

\begin{quote}
\textit{Now we all intuitively know that, more hands on data, more risk; more hands on data potentially, more, higher scale of risk.} (CON2)
\end{quote}

\begin{quote}
\textit{there's two or three vectors that you have to consider us on... how sensitive is it, are we special category data, are we children, are we criminal records,... are we dealing with vulnerable groups of people.} (CON2)
\end{quote}

\subsubsection{Tools and guidance for DPIA}

A useful method to tackle difficulty of running DPIA is for privacy professionals to follow readily available guidelines and templates for DPIA such as NHS templates, ICO template \cite{ICO-template}, CNIL \cite{CNIL}, ENISA \cite{ENISA}, cyber insurance templates or self-designed Excel spreadsheets platforms used by individual organizations.GDPR article 35 also instructs compliance with code of conducts for DPIA and advises data protection board and supervisory authority to provide such guidelines to organizations. Supervisory authorities and member states shall also encourage producing such code of conducts. Associations in charge of organizations and businesses are recommended to prepare the code of conducts and must submit them to supervisory authorities for their approval, registration and publication. In response to this requirement the supervisory authority in the UK itself has published guidelines to DPIA \cite{ICO-DPIA-Guidance}.  However, participants in our focus groups addressed usefulness and being basic of available templates and guidelines from ICO. In general, although some of other available platforms had more positive feed-backs by our interviewees, it was concluded that none of the available guidelines are useful as a unique platform. Some have been reported  to be used only to authorize and credit DPIA tasks. ICO guidelines have been reported to be basic and manual, making it hard to be used.

\begin{quote}
\textit{the limited value of the different pieces of advice that out there, the different guidance documents. There's no one document that you can go and say I'll use this.} (CON1)
\end{quote}

\begin{quote}
\textit{The templates I've come into that are ICO templates, and everything is ICO templates where I am, are so basic they're just killing me because they're so manual.  It's really hard to run with them.} (CON3)
\end{quote}

Measuring the risk level should include measuring both the likelihood and impact of the risk, based on GDPR and Article 29 Data Protection Working Party \cite{EC-Data-protection}. Both GDPR and Article 29 provide data controllers with flexibility to determine which methodology or precise structure to use for DPIA. However, they recommend a sector-specific framework that can be bespoke to the data controller. Common criteria as the general approach to DPIA has been recommended by Article 29 Working Party. Most of our participants mentioned usage of NHS matrix as the measurement tool for risk assessment in DPIA and found it very useful.

\begin{quote}
\textit{we have a really simple table; likelihood against severity and we assess a numerical kind of score into likelihood against severity of impact}  (DPO2)  
\end{quote}

\begin{quote}
\textit{I keep harking back to the NHS ...The latest incident reporting guidance actually has a severity matrix which I think is quite good.  It goes from zero to five but it's not the classic red, green, yellow pattern that everybody's familiar with...it's much better than the normal five by five we're familiar with} (DPO1)
\end{quote}

An uniform platform for DPIA is still not available to practitioners and current guidelines are not easy to be followed or used.

Risks associated to personal data processing have been generally defined in GDPR as high risks to the rights and freedom of natural persons taking into account the nature, scope, context and purpose of processing. GDPR has listed risk assessment activities which among are description of envisaged processing operation and processing purposes, the necessity of processing, and assessment of the risks to the rights and freedoms of data subjects. These  activities are to identify the resources of risks or to say system vulnerabilities. Our focus group discussion, identified number of the associated context to risk identification as processing purpose, vendors, systems, applications, websites, human factors and others. Using depth questionnaires about system context has been identified as an approach to the identification of risks. Some resources such as EPDB or WP29 had been identified as good benchmarks for designing these questioners.However, the need for a predefined database of standardized and codified risks has been recognized as a gap in current DPIA practices. NCSC \cite{NCSC} archives have been identified as one of the valuable resources for design of such a database. 

\begin{quote}
\textit{You go through your data privacy impact assessment again and your legitimate business, your reason, your consent, your legitimate business interests, contractual, whatever, for that purpose.} (CON3)
\end{quote}

\begin{quote}
\textit{There are massive economies of scale when gathering risk intelligence...I'm really just trying to codify and operationalize the kind of deep experience and expertise of the kind of people...so we don't have to plug a specialist into every step of the job  } (CON2)
\end{quote}

\begin{quote}
\textit{Cyber insurance, interesting one...but of course they're guarded because their commercial value is in it} (CON1)
\end{quote}

\subsubsection{Treatment Controls}

The last thing to do in a risk assessment process is to treat the risks identified with different strategies and controls to treat, transfer, ignore or accept the risk . Various types of treatment controls such as human factor, role access, access controls, encryption, pseudonymization etc have been discussed by our practitioners. Their weakness and strength were discussed,  along with the fact that despite all efforts, data incidents may still happen. However it is important to demonstrate you have done your bests, although it wont withdraw data controller's responsibilities in case of breach.The idea to create repositories of high level controls and compliance practices has been discussed in the group as well.

\begin{quote}
\textit{You need a proactive logging system... we call a VIP alert... It's about the role-based access controls. } (CON3)
\end{quote}

\begin{quote}
\textit{if you as a specialist assessment have assessed your privacy, security controls to be within your appetite,..  You've gone as far as you can to get your residual risk to an acceptable level and still be able to do your job, and something still goes wrong,...It doesn't matter if you've secured it well, if you never had permission to have it in the first place, if you didn't have a right lawful basis...But if you've had all that, then from a regulatory sanctions point of view it is highly unlikely that you would end up with fines from an information commissioners' point of view,...it doesn't matter if you've got controls in place, you still have a level of responsibility.  
There's residual risk} (CON2)
\end{quote}

\begin{quote}
\textit{I got a real bee in my bonnet about encryption companies saying, oh we're going to solve your GDPR problem, no you're not...But as soon as it involves credentials, as so much does, you know, guessing, Brute Force, Social Engineering, FAFT, you're not solving or reducing the risk of a lot of crime by encrypting.} (CON2)
\end{quote}

\begin{quote}
\textit{Pseudonymized within the lower risk community, anonymized within the higher risk community.} (CON2)
\end{quote}

\begin{quote}
\textit{There's some ideas going on in that space, with people trying to create repositories of high'ish level controls and practice compliance, that you can be in a position to access via those entities} (CON2)
\end{quote}

Privacy by design (PbD) as one of the main elements in GDPR, was firstly coined by Ann Cavoukian \cite{cavoukian2010privacy} with the aim to integrate privacy as a system requirement into the entire software development life cycle. The requirement for privacy by design was also indirectly indicated in Article 17 of European Data Protection Directive with its meaning being augmented in Recital 46 highlighting the measures to be taken against privacy risks in system design \cite{horodyski20142013}. However, the legal requirements for controllers to have privacy by design in their development has been directly enforced by GDPR, making it one of the essentials of DPIA. Privacy by design was not reported in our focus group as a very well-known task to the practitioners. They were aware of the requirement for privacy by design, however they did not know what it is or how to do it. It was reported to have been mostly practiced or known by public organizations rather than commercial and private companies which have only after the fact privacy. Another part of the discussions proved that although the Privacy by Design was not very well known to the participants, but they were already aware and had practiced some PbD activities such as data flowing before. GDPR has framed all the available privacy practices under a unique and systematic platform, obviously with larger enforcement, which needs more awareness and training.     

\begin{quote}
\textit{But this privacy by design thing, which everybody keeps coming to me and saying what are we supposed to be doing?  You know, what is this privacy by design?  I say oh yes, I know all about that, when inside a little voice is going `Run away!  Run away!'} (CON1)
\end{quote}

\begin{quote}
\textit{So that, I think, has been the big changes since the new data protection law.  The interesting thing is how many of these things have been around for a very, very long time but they've been an elephant in the room, and I think that elephant now is becoming more.} (DPO1)
\end{quote}

\subsubsection{Integration of DPIA with Project Risk Assessment}   
 
Risk assessment has been more highlighted in GDPR as it is having been defined as a risk assessment-based approach. Recital 74 of GDPR has stated obligations to controllers to take into account risks to the rights and freedom of natural persons \cite{EU}. Having privacy risks likelihood and impact being assessed in DPIA, GDPR effectively has incorporated a risk-assessment approach into data protection. However, there is no agreed definition or concept of ``risk'' and its notion in the GDPR. Along with privacy risks, there are other types of risks such as information security, financial, health \ safety, etc, being assessed and considered by organizations using different types of general risk management/ project management or specific to a subject (e.g. information security) approaches. Assuming that integration of DPIA into organizational project risk assessment will make DPIA process more common and used, some attempts had been done in recent years to communize privacy risk assessment with other types of organizational risks \cite{wright2014integrating}. This also has attracted part of our participants conversation in the second study. Generally, they all agreed with an integrated DPIA into the organizational risk assessment, however some believed in separation between privacy and other risk assessment process such as information security although under same umbrella. The accountability and responsibility of risk assessment, privacy risk assessment and information security risk assessment were also an area of doubt as most of participants believed that privacy officers are not qualified risk assessors and based on GDPR, DPOs are the only accountable for the privacy risk assessment. Putting the responsibility on information or general risk assessor, then the need for a close cooperation between these roles is necessary to resolve possible conflicts, which sometimes may not be feasible.  

\begin{quote}
\textit{But to be fair that's not a lot different to the risk management for everything else in most businesses, so you're not going to be out of sync with the rest of their risk management efforts.} (CON1)
\end{quote}

 \begin{quote}
\textit{...been in the corporate world, they also saw the data protection officer should be doing the risk assessment for them and everything else.  So we used to have to push back to them and say it's your process.} (CON3)
\end{quote}

\begin{quote}
\textit{We have the role called senior information risk owner in most public sector organizations, which is what it says on the tin.  They own the organization's information risk and I've had some fairly robust discussions with SIRO and essentially had to say to them look, you know, it is my job as the data protection officer to protect you as an organization but also to protect the data subjects.  So if you as an organization are doing something which I think is completely the pale then obviously I will be challenging you about it.} (DPO1)
\end{quote}

\subsubsection{Training}

The requirement for privacy training and education of organizational staffs as the direct agencies dealing with daily personal data processing of individuals has been addressed in Articles 39, 47 and 70 of GDPR \cite{EU} mostly in form of responsibilities of Data Protection Officers. Various supervisory authorities such as ICO have also published guidelines for training \cite{ICO-training} since after GDPR enforcement making staff privacy training as one of the essentials for GDPR compliance. However, still many concerns have been raised in the second focus groups as for the inconvenience and abstract level of privacy training and education for organization staffs. They have been reported to be almost based on noticeably short questioners and only for commercial objectives to prove GDPR certification or to skip the responsibility in data breach cases.  

\begin{quote}
\textit{I've done my training. It took me ten minutes; I could just click through.} (DPO2)
\end{quote}

\begin{quote}
\textit{you can minimize the impact on your reputation and the finances of your school by pushing it onto the individual, by saying, well we trained that person.} (DPO2)
\end{quote}

Data protection practices such as training are reported  to be basic and only for sake of accountability.

Part of the difficulty looks to be due to low level of skills, un-experience and un-qualified risk assessing practitioners and newly hired privacy practitioners in organizations. Also, the fact that GDPR and its requirements are still a new subject is another obstacle which might be resolved after some years of experience and by more training and education or by development of easy to be used approaches for DPIA. 

\begin{quote}
\textit{They say they don't know what they're doing, and I say believe me, I promise you, in five years' time, this will all come naturally to you.} (DPO2)
\end{quote}

\subsubsection{DPIA Transparency}

Publishing DPIA is not a legal requirement based on GDPR. However, many supervisory authorities such as ICO advise and recommend DPIA publication to demonstrate compliance \cite{ICO-Accountability}, also to establish trust with data subjects and public. This subject had also attracted part of the conversation in the new focus group, where participants agreed on some level of transparency for DPIA, due to the security risks that the whole transparency may develop to personal data processes.   

\begin{quote}
\textit{Yeah you can request it, and they should be published to some extent, but probably not the whole thing, it depends on the nature of the business.  Because there might be elements in the data, in a full data protection impact assessment which allude to security techniques.} (CON1)
\end{quote}

 
 

 
\subsection{Challenges}

\subsubsection{Over-complicated and time-consuming}

The overall position of our participants on the complexity and processing time of DPIA was negative, meaning shorter and easier procedure for DPIA is still in demand. They also stated how it depends on the experience and knowledge of practitioners (risk assessors) and the case itself, how it may become easier and faster to perform after some years of experiences. They also reported on difficulty of running very long and complicated data protection surveys and questioners from the consumers, which should be simplified to a questioner of at most 20 minutes of simple questions. 

\begin{quote}
\textit{and asking all those questions in a way that people can understand before they've done the detailed design [...] if we make it quick and simple we get a high voluntary completion rate.  I've been up to eighty percent with a fifteen minute survey that gets, in aggregate, massive amount of beneficial inherent risk information.} (CON2)   
\end{quote}

\begin{quote}
    \textit{... trying to explain complex concepts sometimes to people who have literally no frame of reference to understand that, ...  You really are having to go from basic, basic principles and that is quite time consuming} (DPO1)
\end{quote}

 \begin{quote}
     \textit{... in five years’ time, this will all come naturally to you.} (DPO2)
 \end{quote}
 
 \begin{quote}
     \textit{You can’t time it no, it could be five minutes, five hours depending on what you’re assessing.} (CON1)
 \end{quote}

 \begin{quote}
     \textit{but that’s no good if I then make it so complicated that the school cannot engage with it at all and they’re just completely bamboozled, and I think what you’ve said about being able to do their part of the work in 20 minutes is really valid.} (DPO2)
 \end{quote}
 
\subsubsection{Compliance with the GDPR}

Our analysis on themes from the second focus group identified compliance with the GDPR as the most challenging issue being discussed by practitioners. It was discussed that most of organizations claiming compliance to GDPR on their websites, are not compliant. It was explained how the process has been taken easy only through a short training course for GDPR for their staffs. It was concluded that a 100\% compliance to GDPR is never achievable, and it can only be assured by a proper action plan including a decent risk assessment-based approach. Therefore, compliance shall be proved by logged evidence such as privacy risk registers, standard certificates and DPIA.  

\begin{quote}
\textit{Can you ever, ever be 100 per cent?  You can be as compliant with action plans and risk registers and everything else, but you can never be 100 per cent compliant but you can be heading towards compliance.} (CON2)
\end{quote}

\begin{quote}
\textit{Unfortunately, so many of vendors that we're using don't have anything anywhere near ISO 27001...IT companies don't have that or they don't have Cyber Essentials} (DPO2)
\end{quote}

The other area of GDPR compliance touched by the practitioners was the difference between regulatory and ethical compliance. Legal or regulatory compliance to laws such as GDPR has been differentiated from ethical compliance. Ethical compliance is to confirm system adherence to ethical and moral principles such as trust, fairness, user's expectations, and rights, which is mostly an absent in most areas and only have been taken voluntarily by some organizations  with no general approach to it. Therefore, it is difficult and same time vague for privacy practitioners how to achieve ethical compliance in their regulatory tools such as DPIA.  

\begin{quote}
\textit{Whilst there's a parameter for defining legal compliance, and you can clearly see what you've done that is actually GDPR compliance, what can you say you've done that is ethical?  And you've got all the ethical principles, and all of that, but how do you measure that, who is the ethics police for that?} (ACA2)
\end{quote}

\begin{quote}
\textit{They would expect you to be ethical as well, and I really struggle with trying to be ethical...when I don't even know what is ethical, but I know there's legal compliance} (ACA2)
\end{quote}

\subsubsection{Privacy Roles}

GDPR has introduced and obligated a new role for privacy accountability in organizations known as Data Protection Officer. However, the responsibilities of this role, and distinguish between this role and other privacy and risk relevant positions and who should do DPIA in organizations was a concern of our practitioners in the new study. 

\begin{quote}
\textit{Risk assessment is a corporate responsibility and not the responsibility of the data protection officer of the information governance leaders, they may have been formerly.} (DPO1)
\end{quote}

\begin{quote}
\textit{There's still a wide responsibility with our non-data protection professional colleagues who think that responsibility for anything data protection related devolves to you.  It's for you to do the DPIA.  It's for you to deal with subject access requests, right policies, etc.} (CON2)
\end{quote}

\section{Discussion and Comparison of before and after GDPR Studies}\label{sec:Discussion} 

Based on our thematic analysis on our second focus group study and its comparison to our outcome from the first study before the GDPR date, we still could find some matters outstanding or being newly discovered in privacy practitioners' activities.
In summary, the three challenges mentioned in our previous study as DPIA being over-complicated and time-consuming,  being guts and fast moving of technology and its privacy issues have still remained outstanding after GDPR. 
Despite GDPR rules and requirements, improve in public awareness of data protection concepts and some un-arranged skills for DPIA from previous experiences, there is still needs for a systematic procedure for DPIA along with number of new issues that has arisen after GDPR. The main challenge remains as a need for quantification DPIA with consideration and integration of legal compliance with ethical compliance which all privacy and moral rights of data subjects are taken care of.

\section{Conclusion}\label{sec:Conclusion}
Risk assessment and risk management are established as scientific field with long history in research which has also contribution in practical implementation of risk analysis \cite{aven2016risk}. Because many open challenges faced by privacy professionals are common among risk assessors from other domains such as information security or finance, a systematic study on risk assessment in general can help to integrate the findings and to provide a scientific framework for Data Protection Risk Assessment. The result can be applied to practical activities in this domain, therefore shifting from an experience-based privacy risk assessment to a scientific and systematic one. This should solve the issues identified here about not availability of an unique, non-manual, quantifying and easy to be used template for DPIA which considers all human factors on measurement of likelihood and impact on data subjects. Necessity for an integrated approach of DPIA with project risk assessment is another strong motivation for this move. Although research and works in this field had been initiated before \cite{wright2014integrating}, our recent focus group after GDPR implementation evidenced that the matter has remained, and a solution is vital.  

\appendix
\section{Focus group questions}
\label{appendix:questions}
\begin{itemize}
    \item How do you assess privacy risk?
    \item How do you assess the likelihood of privacy risks?
    \item How do you assess the impact of privacy risks?
    \item How does the scale of privacy risks influence the impact?
    \item How do you assess the scale of privacy risks?
    \item How does the sensitivity of data influence the impact?
    \item How do you assess the sensitivity of data?
    \item How do user expectations influence the impact?
    \item How do you assess user expectations?
    \item How does the harm caused influence the impact?
    \item How do you assess the harm caused by privacy violations?
    \item What other factors influence the impact of privacy risks?
    \item What gaps do you see in privacy risk assessment?
    \item How do you rate the quality of publicly available guidance, e.g. from the ICO or the Article 29 Data Protection Working Party?
\end{itemize}

\bibliography{references}

\begin{thebibliography}{10}
\providecommand \doibase [0]{http://dx.doi.org/}%

\bibitem{warren1890right}
Warren SD, Brandeis LD. The {{Right}} to {{Privacy}}. {\it Harvard Law Review}
  1890\string; 4(5)\string: 193--220.
\newblock \href {\doibase 10.2307/1321160} {doi: 10.2307/1321160}

\bibitem{birnhack2008eu}
Birnhack MD. The {{EU Data Protection Directive}}: {{An}} Engine of a Global
  Regime. {\it Computer Law \& Security Review} 2008\string; 24(6)\string:
  508--520.
\newblock \href {\doibase 10.1016/j.clsr.2008.09.001} {doi:
  10.1016/j.clsr.2008.09.001}

\bibitem{clarke1997introduction}
Clarke R. Introduction to dataveillance and information privacy and definitions
  of terms. {\it www. anu. edu. au/people/Roger. Clarke/DV/Intro. html} 1997.

\bibitem{ICO}
Information Comissioner's Office~UK I. Privcy Impact Assessment Handbook,
  Version 2.0.
  https://www.huntonprivacyblog.com/wp-content/uploads/sites/28/2013/09/PIAhandbookV2.pdf;
  2013.

\bibitem{ICO-GDPR}
Information Commissioner's~Office I. Guide to the General Data Protection
  Regulation (GDPR) Data protection.
  https://ico.org.uk/media/for-organisations/guide-to-the-general-data-protection-regulation-gdpr-1-0.pdf;
  2018.

\bibitem{wright2014privacy}
Wright D, Raab C. Privacy principles, risks and harms. {\it International
  Review of Law, Computers \& Technology} 2014\string; 28(3)\string: 277--298.

\bibitem{ferra2020challenges}
Ferra F, Wagner I, Boiten E, Hadlington L, Psychoula I, Snape R. Challenges in
  assessing privacy impact: Tales from the front lines. {\it Security and
  Privacy} 2020\string; 3(2)\string: e101.

\bibitem{binns2017data}
Binns R. Data protection impact assessments: a meta-regulatory approach. {\it
  International Data Privacy Law} 2017\string; 7(1)\string: 22--35.

\bibitem{de2012human}
De~Hert P. A human rights perspective on privacy and data protection impact
  assessments. In: Springer.  2012 (pp. 33--76).

\bibitem{wright2014integrating}
Wright D, Wadhwa K, Lagazio M, Raab C, Charikane E. Integrating privacy impact
  assessment in risk management. {\it International Data Privacy Law}
  2014\string; 4(2)\string: 155--170.

\bibitem{horodyski20142013}
Horodyski D. 2013 OECD Guidelines on the Protection of Privacy and Transborder
  Flows of Personal Data as an Example of Recent Trends in Personal Data
  Protection. {\it Collective human rights} 2014\string: 255.

\bibitem{bennett1990formation}
Bennett CJ. The formation Of a Canadian privacy policy: the art and craft of
  lesson-drawing. {\it Canadian Public Administration} 1990\string;
  33(4)\string: 551--570.

\bibitem{bu2017cross}
Bu-Pasha S. Cross-border issues under EU data protection law with regards to
  personal data protection. {\it Information \& Communications Technology Law}
  2017\string; 26(3)\string: 213--228.

\bibitem{hustinx2013eu}
Hustinx P. EU data protection law: The review of directive 95/46/EC and the
  proposed general data protection regulation. {\it Collected courses of the
  European University Institute’s Academy of European Law, 24th Session on
  European Union Law} 2013\string: 1--12.

\bibitem{robinson2009review}
Robinson N, Graux H, Botterman M, Valeri L. Review of the European data
  protection directive. {\it Rand Europe} 2009.

\bibitem{tikkinen2018eu}
Tikkinen-Piri C, Rohunen A, Markkula J. EU General Data Protection Regulation:
  Changes and implications for personal data collecting companies. {\it
  Computer Law \& Security Review} 2018\string; 34(1)\string: 134--153.

\bibitem{sirur2018we}
Sirur S, Nurse JR, Webb H. Are {{We There Yet}}? {{Understanding}} the
  {{Challenges Faced}} in {{Complying}} with the {{General Data Protection
  Regulation}} ({{GDPR}}). In: {Association for Computing Machinery}. ; 2018;
  {Toronto, Canada}\string: 88--95

\bibitem{agarwal2016towards}
Agarwal S. Towards dealing with GDPR uncertainty. {\it IFIP Summer School}
  2016.

\bibitem{tankard2016gdpr}
Tankard C. What the GDPR means for businesses. {\it Network Security}
  2016\string; 2016(6)\string: 5--8.

\bibitem{team2020eu}
Team IGP. {\it Eu general data protection regulation (gdpr)--an implementation
  and compliance guide}.
\newblock IT Governance Ltd .
\newblock 2020.

\bibitem{raab2020information}
Raab CD. Information privacy, impact assessment, and the place of ethics. {\it
  Computer Law \& Security Review} 2020\string; 37\string: 105404.

\bibitem{weaver2013corporate}
Weaver GR, Trevi{\~n}o LK, Cochran PL. Corporate ethics practices in the
  mid-1990s: An empirical study of the Fortune 1000. In: Springer.  2013 (pp.
  625--640).

\bibitem{CNIL}
Commission Nationale Informatique \&~Libertes C. Privcy Impact Assessment (PIA)
  Methodology.
  https://www.cnil.fr/sites/default/files/atoms/files/cnil-pia-1-en-methodology.pdf;
  2018.

\bibitem{wright2013comparative}
Wright D, Finn R, Rodrigues R. A comparative analysis of privacy impact
  assessment in six countries. {\it Journal of Contemporary European Research}
  2013\string; 9(1).

\bibitem{warren2012privacy}
Warren A, Charlesworth A. Privacy impact assessment in the UK. In: Springer.
  2012 (pp. 205--224).

\bibitem{wright2013making}
Wright D. Making privacy impact assessment more effective. {\it The Information
  Society} 2013\string; 29(5)\string: 307--315.

\bibitem{EU}
European~Union E. General Data Protection Regulation. https://gdprinfo.eu/;
  2016.

\bibitem{van2017privacy}
Puijenbroek vJ, Hoepman JH. Privacy impact assessments in practice: Outcome of
  a descriptive field research in the Netherlands.  2017.

\bibitem{wagner2018privacy}
Wagner I, Boiten E. Privacy {{Risk Assessment}}: From {{Art}} to {{Science}},
  by {{Metrics}}. In: . LNCS 11025. {Springer}. ; 2018; {Barcelona,
  Spain}\string: 225--241

\bibitem{leech2007array}
Leech NL, Onwuegbuzie AJ. An array of qualitative data analysis tools: A call
  for data analysis triangulation.. {\it School psychology quarterly}
  2007\string; 22(4)\string: 557.

\bibitem{onwuegbuzie2009qualitative}
Onwuegbuzie AJ, Dickinson WB, Leech NL, Zoran AG. A qualitative framework for
  collecting and analyzing data in focus group research. {\it International
  journal of qualitative methods} 2009\string; 8(3)\string: 1--21.

\bibitem{goddard2017eu}
Goddard M. The EU General Data Protection Regulation (GDPR): European
  regulation that has a global impact. {\it International Journal of Market
  Research} 2017\string; 59(6)\string: 703--705.

\bibitem{kuner2012european}
Kuner C. The European Commission's proposed data protection regulation: A
  copernican revolution in European data protection law. {\it Bloomberg BNA
  Privacy and Security Law Report (2012) February} 2012\string; 6(2012)\string:
  1--15.

\bibitem{aven2018society}
Aven T, Ben-Haim Y, Boje~Andersen H, et al. Society for risk analysis glossary.
  {\it Society for Risk Analysis, August} 2018.

\bibitem{OECD}
Economic Co-operation fO, Development O. What is Impact Assessment?).
  https://www.oecd.org/sti/inno/What-is-impact-assessment-OECDImpact.pdf;
  2014.

\bibitem{meis2015supporting}
Meis R, Heisel M. Supporting privacy impact assessments using problem-based
  privacy analysis. In: Springer. ; 2015\string: 79--98.

\bibitem{schweizerische2013information}
Schweizerische S. Information technology-Security techniques-Information
  security management systems-Requirements. {\it ISO/IEC International
  Standards Organization} 2013.

\bibitem{EC-Data-protection}
European~Commission WP. Guidance on data protection risk assessment (DPIA) and
  determining if the processing is "likely to result in high risks" for the
  purposes of regulation 2016/679, wp248.rev01.
  \url{https://ec.europa.eu/newsroom/just/document.cfm?doc_id=47711};  2017.

\bibitem{ICO-template}
Information Commission~Office I. Sample DPIA Template.
  https://ico.org.uk/media/about-the-ico/consultations/2258461/dpia-template-v04-post-comms-review-20180308.pdf;
  2018.

\bibitem{ENISA}
Cybersecurity E.~U. A.~fE. tool for security of personal data processing.
  https://www.enisa.europa.eu/risk-level-tool;  2017.

\bibitem{ICO-DPIA-Guidance}
Information Commissioner's~Office I. Data protection impact assessments.
  https://ico.org.uk/for-organisations/guide-to-data-protection/guide-to-the-general-data-protection-regulation-gdpr/;
  2019.

\bibitem{NCSC}
National Cyber Security~Center N. Reports \& advisories.
  https://www.ncsc.gov.uk/section/keep-up-to-date/reports-advisories;  2022.

\bibitem{cavoukian2010privacy}
Cavoukian A, Taylor S, Abrams ME. Privacy by Design: essential for
  organizational accountability and strong business practices. {\it Identity in
  the Information Society} 2010\string; 3(2)\string: 405--413.

\bibitem{ICO-training}
Information Commissioner's~Office I. Training and awareness.
  https://ico.org.uk/for-organisations/accountability-framework/training-and-awareness/;
  2020.

\bibitem{ICO-Accountability}
Information Commissioner's~Office I. Accountability and governance Data
  Protection impact Assessments (DPIAs).
  https://ico.org.uk/media/for-organisations/guide-to-the-general-data-protection-regulation-gdpr/data-protection-impact-assessments-dpias-1-0.pdf;
  2018.

\bibitem{aven2016risk}
Aven T. Risk assessment and risk management: Review of recent advances on their
  foundation. {\it European Journal of Operational Research} 2016\string;
  253(1)\string: 1--13.

\end{thebibliography}

\end{document}